# Performance Comparison of a proposed Vector Tracking architecture versus the Scalar configuration for a L1/E1 GPS/Galileo receiver

Enik Shytermeja, Axel Garcia-Pena, Olivier Julien
TELECOM/SIGNAV Lab, ENAC, Toulouse, France
Email : shytermeja, garcia-pena, ojulien@recherche.enac

*Abstract*— In urban environments, standalone GNSS receivers can be strongly affected to the point of not being able to provide a position accuracy suitable for use in vehicular applications. In this paper, a vector delay/frequency-locked loop (VDFLL) architecture for a dual constellation L1/E1 GPS/Galileo receiver is proposed. In this implementation, the individual DLLs and FLLs of each tracked satellite are replaced with an Extended Kalman filter (EKF), responsible for both estimating the user's position, velocity and clock bias and closing the code/carrier updates for each GPS L1 and Galileo E1 tracking channels. In this work, a detailed performance comparison between the scalar tracking and VDFLL configuration is assessed under signal outages and significant power drops conditions that are simulated in different satellite channels. Contrary to the conventional tracking, the L1/E1 VDFLL loop is able to recover the frequency and code-delay estimation at the end of the simulated outages without the requirement of signal reacquisition process.

## I. INTRODUCTION

In the last decade, Global Navigation Satellites Systems (GNSS) have gained a significant position in the development of Urban Navigation applications and associated services. A major concern of the constant growth of GNSS-based urban applications is related to the quality of the positioning service, expressed in terms of accuracy, availability and continuity of service but also of integrity provision, ensuring that the application requirements are met [1]. In urban environments, standalone GNSS receiver architectures can be strongly affected to the point of not being able to provide a position accuracy suitable for use in vehicular applications. Specifically, the reception of GNSS signals is affected by the surrounding objects, such as high buildings, trees, lampposts and so on, which can block, shadow, reflect and diffract the received signal. As a result, two significant signal distortions are generated.

On one hand, the reception of reflected or diffracted GNSS LOS echoes in addition to the direct LOS signal generates the phenomenon known as multipath. Multipath echoes represent one of the most detrimental positioning error sources in urban canyons. In fact, the reception of echoes distorts the ideal correlation function and leads to a degradation of the signal code and carrier estimations accuracy up to a loss of lock of the code and carrier tracking loops. Consequently, the pseudo-range and Doppler measurements are degraded.

On the other hand, the total or partial obstruction of the GNSS LOS by the urban environment obstacles causes GNSS LOS blockage or GNSS LOS shadowing phenomena. The reception of Non-LOS (NLOS) signals can then introduce a bias on the pseudo-range measurements if only NLOS satellites are tracked. This bias can be very important as it is representative of the extra-path travelled by the NLOS signal compared to the theoretical LOS signal. The LOS shadowing can also decrease the LOS signal $C/N_0$ and thus makes the signal more vulnerable to the multipath effect.

Finally, the resulting degraded measurements cause the navigation processor to compute an inaccurate position solution or even to be unable to compute one in the case of few available measurements. Therefore, advanced GNSS signal processing techniques must be implemented in order to improve the navigation solution performance in urban environments.

Conventional GNSS receivers basically consist of two units such as, the signal processing module that performs the signal acquisition and tracking tasks for both the code delay and carrier frequency/phase offset and secondly, the navigation module providing the user navigation solution and clock terms estimation. Moreover, in scalar tracking configuration in the presence of weak signals or significant signal power drops, loss of lock of the affected satellite occurs and therefore, its estimated pseudoranges are not passed to the navigation processor due to their lack of accuracy.

A promising approach for reducing the effect of multipath interference and NLOS reception is vector tracking (VT), first introduced in [2] where the signal tracking and navigation solution tasks are accomplished by the central navigation filter. In comparison to conventional or scalar tracking (ST), where each visible satellite channel is being tracked individually and independently, VT performs a joint signal tracking of all the satellite channels. Vector tracking exploits the knowledge of the estimated receiver's position and velocity to control the receiver's tracking feedback. In [2], the Vector Delay Lock Loop (VDLL) architecture is explained in details, for which the navigation filter replaces the delay lock loops (DLLs) with an Extended Kalman filter (EKF). In this configuration, the navigation solution drives the code Numerical Control Oscillator (NCOs) of each tracking channel while the carrier frequency/phase estimation is still achieved scalarly by the Frequency or Phase Lock Loops (FLLs or PLLs). Vector DLL (VDLL) tracking performance of the GPS L1 signal in weak signal-to-noise ratio (SNR) environment and robustness against signal interference and attenuation has been demonstrated in [3], [4] and [5].

The objective of this paper is to assess the performance of the Vector Delay Frequency Lock Loop (VDFLL) architecture, seen as a combination of the VDLL and VFLL loops, in signal-constrained environment. From the navigation point of view, VDFLL represents a concrete application of information fusion, since all the tracking channels Numerical Control Oscillators (NCOs) are controlled by the same navigation solution filter.

In this paper, a dual constellation GPS + Galileo single frequency L1/E1 VDFLL architecture is presented since this type of receiver can significantly improve the availability of a navigation solution in urban canyons and heavily shadowed areas: an increased satellite in-view availability is directly translated in a higher measurement redundancy and improved position reliability. A detailed performance comparison between the scalar tracking and VDFLL configuration in terms of position and code/carrier tracking accuracies is assessed in a simplified urban environment, where signal outages and significant power drops are simulated in different satellite channels.

## II. PROPOSED L1/E1 VDFLL ARCHITECTURE

The proposed VDFLL architecture comprises three sub-modules: the code/carrier tracking loops including the DLL/FLL discriminators, the EKF navigation filter and the code/carrier NCOs update. In this work, we present the dual constellation single frequency band L1/E1 VDFLL architecture, wherein the code (DLL) and frequency (FLL) tracking loops are coupled through the navigation solution computed by the central extended Kalman filter (EKF). The detailed architecture of the proposed L1/E1 VDFLL configuration is sketched in Fig.1.

Kalman filter estimation equations fall into two groups:

- *Time update (Prediction)* equations, performing the forward projection in time of the current vector state $X_k^+$ and the state error covariance matrix $P_k^+$ "*a priori*" estimates for the next time epoch, $X_{k+1}^-$ and $P_{k+1}^-$, where $k$ indicates the current time epoch;

- *Measurement update (Correction)* equations, responsible for the feedback that is achieved by feeding the current epoch measurement vector, denoted as $z_{input}$ into the *a priori* estimate, $X_{k+1}^-$ and $P_{k+1}^-$, to obtain an improved *a posteriori* estimate, $X_{k+1}^+$ and $P_{k+1}^+$.

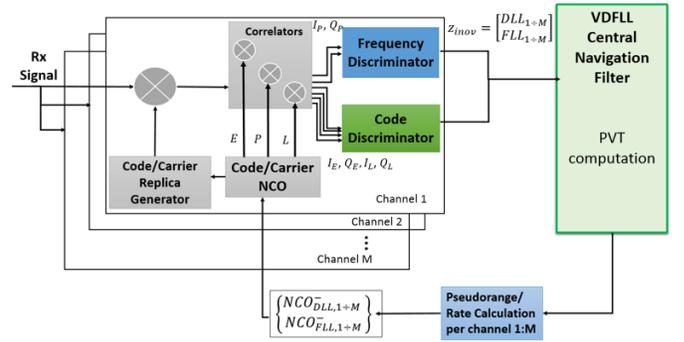

**Fig. 1.** The proposed L1/E1 VDFLL architecture.

### A. EKF STATE MODEL

The chosen state vector model in our EKF navigation filter implementation is the Position and Velocity (PV) representation, containing the following states:

$$X_k = [x \; \dot{x} \; y \; \dot{y} \; z \; \dot{z} \; c \cdot t_{GPS-clk} \; c \cdot t_{GAL-clk} \; c \cdot \dot{t}_{clk}]_k^T, \quad (1)$$

being a 9x1 absolute state vector, containing both the receiver's position vector $[x(k), y(k), z(k)]^T$ and the receiver's velocity vector $[\dot{x}(k), \dot{y}(k), \dot{z}(k)]^T$ in ECEF coordinates; the receiver's clock dynamics comprising the receiver clock bias w.r.t the GPS and Galileo time and drift components $[c \cdot t_{GPS-clk} \; c \cdot t_{GAL-clk} \; c \cdot \dot{t}_{clk}]^T$, where $c$ is the speed of light and therefore the clock biases and drift are expressed in unit of [m] and [m/s], respectively. Assuming that satellite clock biases are perfectly corrected, the bias increase (and thus the rate) only depends on the receiver's clock. Therefore, there is the same clock drift for both constellation.

The system model of the EKF filter in the continuous time domain may be expressed as:

$$\dot{X}(t) = \varphi \cdot X(t) + B \cdot w(t), \quad (2a)$$

Where $\dot{X}(t)$ denotes the derivative of the state vector $X(t)$, $w(t)$ is the centered gaussian white noise affecting the state vector, $\varphi$ is the system matrix and $B$ is the colored noise transition matrix, both in the continuous time domain.

Passing to the discrete time domain, the system or dynamic model of the VDFLL navigation filter can be detailed as follows:

$$X_k = \Phi \cdot X_{k-1} + w_k, \quad (2b)$$

where: $X_k$ denotes the state vector forward projection from the $k-1^{th}$ to the $k^{th}$ time epoch and $\Phi$ represents the dynamics of the user platform and clock, expressed as follows:

$$\Phi = \begin{bmatrix} C & 0_{2\times 2} & 0_{2\times 2} & 0_{2\times 3} \\ 0_{2\times 2} & C & 0_{2\times 2} & 0_{2\times 3} \\ 0_{2\times 2} & 0_{2\times 2} & C & 0_{2\times 3} \\ 0_{3\times 2} & 0_{3\times 2} & 0_{3\times 2} & C_{clk} \end{bmatrix}_{9\times 9}, \quad (3)$$

where:

$$C = \begin{bmatrix} 1 & \Delta T \\ 0 & 1 \end{bmatrix} \text{ and } C_{clk} = \begin{bmatrix} 1 & 0 & T \\ 0 & 1 & T \\ 0 & 0 & 1 \end{bmatrix} \quad (4)$$

and $\Delta T$ denotes the time interval between two consecutive estimations, representing the measurement update time of the central filter.

The discrete process noise vector $w_k$ is modeled as a white Gaussian noise vector with zero mean and discrete covariance matrix $Q_k$. The process noise $w_k$ comes from two sources namely, the user dynamic noise $[w_x, w_{\dot{x}}, w_y, w_{\dot{y}}, w_z, w_{\dot{z}}]$ (constituted by the user's position and velocity terms) and the receiver's clock noise (local oscillator NCO noise) $[w_b, w_d]$, grouped in a single vector representation as:

$$w_k = [w_x\, w_{\dot{x}}\, w_y\, w_{\dot{y}}\, w_z\, w_{\dot{z}}\, w_{b-GPS}\, w_{b-GAL}\, w_d]_k^T, \quad (5)$$

In Kalman filtering, the process and the measurement noise covariance matrices are very crucial parameters that significantly affect the performance of the filter. Therefore, an accurate tuning is required to fasten the EKF estimation convergence toward the true user state. The discrete-time process noise covariance matrix $Q_k$ is generated from the continuous-domain process noise $Q$ matrix that represents the uncertainty of the user's dynamics. It is modelled based on the influence of five process noise power spectral densities (PSDs) as:

$$Q = E\{w \cdot w^T\}$$
$$= \begin{bmatrix} \sigma_{\dot{x}}^2 & 0 & 0 & 0 & 0 & 0 \\ 0 & \sigma_{\dot{y}}^2 & 0 & 0 & 0 & 0 \\ 0 & 0 & \sigma_{\dot{z}}^2 & 0 & 0 & 0 \\ 0 & 0 & 0 & \sigma_{b-GPS}^2 & 0 & 0 \\ 0 & 0 & 0 & 0 & \sigma_{b-GAL}^2 & 0 \\ 0 & 0 & 0 & 0 & 0 & \sigma_d^2 \end{bmatrix} \quad (6)$$

Based on their nature, the five tuning factors of process noise $Q$ covariance matrix can be grouped in two main categories, such as:

- *User's dynamics sensitive*: including the velocity error variance terms along the ECEF axes $(\sigma_{\dot{x}}^2, \sigma_{\dot{y}}^2, \sigma_{\dot{z}}^2)$ that will be projected in the position domain through the state transition matrix $\Phi$ and the coloured noise transition matrix $B$ from Eq. (2a).

- *Receiver's oscillator noise PSDs*: including the oscillator's phase noise PSD, $\sigma_b$, and the oscillator's frequency noise PSD, $\sigma_d$, which by themselves depend on the Allan variance parameters $h_0$ and $h_{-2}$.

The process noise covariance matrix $Q_k = \text{diag}\,[Q_{x,k}, Q_{y,k}, Q_{z,k}, Q_{c,k}]$ in the discrete domain per each entry can be expressed as:

$$Q_{x,k} = \int_{t_{k-1}}^{t_{k-1}+\Delta T} \Phi_x(T) \cdot Q_x \cdot \Phi_x^T(T) \quad (7)$$

where $Q_x$ represents the process noise covariance matrix in the continuous time domain for the user's position and velocity along the $x$ axes. Thus, the user's dynamics process noise discretization for the position- and velocity- states along the x-axes is computed as:

$$Q_{x,k} = \int_{t_{k-1}}^{t_{k-1}+\Delta T} \begin{bmatrix} 1 & \Delta T \\ 0 & 1 \end{bmatrix} \cdot \begin{bmatrix} 0 & 0 \\ 0 & \sigma_{\dot{x}}^2 \end{bmatrix} \cdot \begin{bmatrix} 1 & 0 \\ \Delta T & 1 \end{bmatrix} \quad (8)$$

Finally:

$$Q_{x,k} = \sigma_{\dot{x}}^2 \cdot \begin{bmatrix} \Delta T^3/3 & \Delta T^2/2 \\ \Delta T^2/2 & \Delta T \end{bmatrix} \quad (9)$$

Similarly, the same logic is applied to obtain the discrete-time process noise covariance matrix for the y- and z-axes user's position projections:

$$Q_{y,k} = \sigma_{\dot{y}}^2 \cdot \begin{bmatrix} \Delta T^3/3 & \Delta T^2/2 \\ \Delta T^2/2 & \Delta T \end{bmatrix}, \quad (10)$$

and,

$$Q_{z,k} = \sigma_{\dot{z}}^2 \cdot \begin{bmatrix} \Delta T^3/3 & \Delta T^2/2 \\ \Delta T^2/2 & \Delta T \end{bmatrix} \quad (11)$$

The receiver's clock noise covariance matrix is computed as:

$$Q_{c,k} = \begin{bmatrix} a_{GPS} & 0 & b \\ 0 & a_{GAL} & b \\ b & b & c \end{bmatrix}, \quad (12)$$

where:

$$\begin{aligned} a_{GPS} &= \sigma_{b-GPS}^2 \cdot \Delta T + \sigma_d^2 \cdot \frac{\Delta T^3}{3} \\ a_{GAL} &= \sigma_{b-GAL}^2 \cdot \Delta T + \sigma_d^2 \cdot \frac{\Delta T^3}{3} \\ b &= \sigma_d^2 \cdot \frac{\Delta T^2}{2} \\ c &= \sigma_d^2 \cdot \Delta T \end{aligned} \quad (13)$$

### B. EKF OBSERVATION MODEL

The non-linear relation between the state and the measurement vector is expressed as follows:

$$z_k = h(x_k) + v_k, \quad (14)$$

where $h$ is the *non-linear* function relating the measurement $z_k$ to the state $X_k$ and $v_k$ is *the measurement noise vector* that is modelled as a zero-mean uncorrelated Gaussian noise process and independent to the process noise $w_k$. The measurement vector $z_k$ comprises the pseudoranges $\rho_j$ and Doppler measurements $\dot{\rho}_j$, output from the code/carrier tracking process for the $j = 1 \div M$ L1/E1 tracking channels:

$$z_k = \left[ \left( \rho_1, \rho_2, \cdots, \rho_j \right) : \left( \dot{\rho}_1, \dot{\rho}_2, \cdots, \dot{\rho}_j \right) \right]_k \quad (15)$$

In the Cartesian ECEF-frame implementation, the pseudoranges $\rho_{j,k}$ per each tracked satellite $j$ are computed as:

$$\rho_{j,k} = \sqrt[2]{(x_{sat\,j,k} - X_k(1))^2 + (y_{sat\,j,K} - X_k(3))^2 \cdots} \\ \sqrt[2]{\cdots + (z_{sat\,j,k} - X_k(5))^2} + X_k(7) \;(or\;X_k(8)) \\ + n_{\rho_{j,k}} \quad (16)$$

While the remaining M-entries of the measurement vector, constituted by the Doppler measurements, are computed as:

$$\dot{\rho}_{j,k} = \left( \dot{x}_{sat\,j,k} - X_k(2) \right) \cdot a_{x,j} + \cdots \\ \cdots + \left( \dot{y}_{sat\,j,k} - X_k(4) \right) \cdot a_{y,j} + \cdots \\ \cdots + \left( \dot{z}_{sat\,j,k} - X_k(6) \right) \cdot a_{z,j} + X_k(9) + + n_{\dot{\rho}_{j,k}} \quad (17)$$

Where $(a_{x,j}, a_{y,j}, a_{z,j})$ the a-terms denote the line-of-sight (LOS) unit vectors from the receiver to the $j^{th}$ satellite along the X, Y and Z axes and $(n_{\rho_{j,k}}, n_{\dot{\rho}_{j,k}})$ denote the zero-mean Gaussian-distributed noise affecting the pseudorange and Doppler measurements, respectively.

The measurement noise vector $v_k$ is modelled as a zero-mean uncorrelated Gaussian noise process and independent to the process noise $w_k$:

$$E[v_k] = 0 \quad (18)$$

$$E[v_k \cdot w_l^T] = 0 \quad (19)$$

$$E[v_k \cdot v_l^T] = \boldsymbol{R_k} \cdot \delta_{kl}, \text{for all } k \text{ and } n \quad (20)$$

where $\delta_{kl}$ denotes the Kronecker's delta, and $\boldsymbol{R_k}$ is the *measurement noise covariance matrix*.

In our vector tracking algorithm, an Early Minus Late Power (EMLP) discriminator has been chosen for both the GPS BPSK and Galileo E1 BOC (1,1) channels. The DLL tracking error variance in presence of thermal noise and in the open-loop configuration, for both GPS L1 and Galileo E1 channels is computed as [8]:

$$\sigma^2_{EMLP-open,j} = \left( \frac{c}{f_{code}} \right)^2 \cdot \left( \frac{C_s}{4 \cdot \alpha \cdot C/N_{0\,est,j} \cdot T_{DLL}} \right) \\ \cdot \left( 1 + \frac{2}{(2 - \alpha \cdot C_s) \cdot C/N_{0\,est,j} \cdot T_{DLL}} \right) (m^2) \quad (21)$$

The FLL performs the Doppler frequency tracking of the incoming signal that is dominated by the satellite-to-receiver motion and the user clock drift. The FLL tracking error variance of the Decision-Directed cross-product (DDCP) discriminator in the open-loop configuration is given by:

$$\sigma^2_{FLL-open,j} = \left( \frac{c}{f_{carr}} \right)^2 \cdot \left[ \frac{1}{\left[ \frac{C}{N_{0\,est,j}} \cdot T_{FLL}^3 \right]} \right] (m^2/s^2), \quad (22)$$

where $T_{DLL}$ and $T_{FLL}$ denote the code and carrier filter integration interval equal to 20 ms; $C_s$ is the code chip spacing (0.5 chips for GPS L1 and 0.2 chips for Gal E1 BOC (1,1); $\alpha$ is a coefficient reflecting the sharpness of the code autocorrelation function (1 for BPSK(1) and 3 for BOC (1,1)); $C/N_{0\,est,j}$ refers to the estimated carrier-to-noise ratio from the tracking loop of the incoming signal from the $j-th$ tracking channel and $f_{code} = 1.023\;Mhz$ and $f_{carr} = 1.57542\;GHz$ denote the L1/E1 code chipping rate and carrier frequency, respectively.

Taking into account Eq. (21) - (22), the measurement noise covariance matrix has in the main diagonal the following entries:

$$R_{jj} = \begin{cases} \sigma^2_{EMLP-open,j} & for \quad j = 1 \cdots M \\ \sigma^2_{FLL-open,j} & for \quad j = 1 \cdots M \end{cases} \quad (23)$$

where the first entry refers to the pseudorange error variance terms for the tracked GPS and Galileo satellites, while the second one is a common term for the pseudorange rate error variance for all tracked satellites.

The EKF time prediction and measurement correction models, implemented in this work and illustrated in Fig.2, will be formulated in details in the following section.

### III. VDFLL ESTIMATION WORKFLOW

Following the VDFLL estimation workflow of Fig. 2, the successive step after the state propagation or prediction, is the computation of the Kalman gain in Step 2.1. For this matter, the measurement prediction $z_k$ and observation matrix $H_k$ shall be calculated. Afterwards, the state vector update is computed from the measurement innovation vector input to the EKF navigation filter, which comprises the code and carrier discriminator outputs from the tracking loops. Finally, the code and carrier NCO update, computed from the EKF filter prediction states, closes the feedback loop to the tracking module and will be given in details in section D.

#### A. MEASUREMENT PREDICTION

The predicted measurement vector $z_k^-$ consists of two entries per satellite tracking channel, in specifics the predicted pseudorange $\rho_{j,k}^-$ and pseudorange rates $\dot{\rho}_{j,k}^-$:

$$z_k^- = [\,\rho_{1,k}^-, \cdots \rho_{M,k}^-, \dot{\rho}_{1,k}^-, \cdots \dot{\rho}_{M,k}^-\,]_{2M \times 1}^T, \quad (24)$$

where $M$ denotes the total nr of tracked GPS + Galileo satellites in the current measurement epoch $k$.

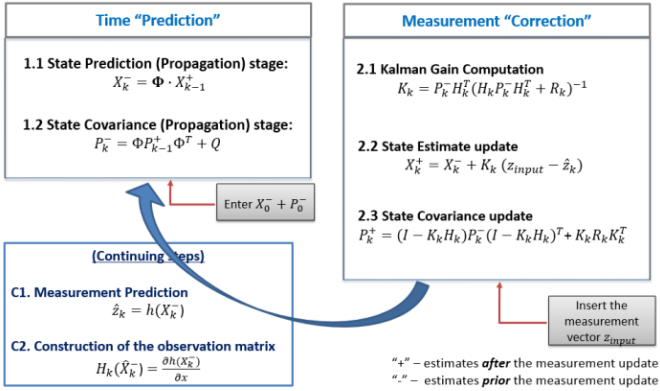

**Fig. 2.** Flowchart of the EKF estimation process

In the Cartesian ECEF-frame implementation, the predicted satellite-user ranges $R_j$ per each tracked satellite $j$ are furtherly computed as:

$$R^-_{j,k} = \sqrt[2]{(x_{sat\,j,K} - X_k^-(1))^2 + (y_{sat\,j,k} - X_k^-(3))^2 \cdots} \sqrt[2]{\cdots + (z_{sat\,j,k} - X_k^-(5))^2} \quad (25)$$

The predicted pseudorange measurement $\rho^-_{j,k}$ can be obtained by adding to the predicted distance $R^-_{j,k}$, the EKF clock bias estimation $x_k^-(7)$:

$$\rho^-_{j-GPS,k} = R^-_{j,k} + X_k^-(7)$$
$$\rho^-_{j-GAL,k} = R^-_{j,k} + X_k^-(8) \quad (26)$$

where, the first expression denotes the predicted pseudoranges from the GPS satellites comprising the predicted user clock bias w.r.t to GPS time; while the former relation is linked to the Galileo-related predicted ranges.
Similarly, the predicted pseudorange rate $\dot{\rho}^-_{j,k}$ can be computed as:

$$\dot{\rho}^-_{j,k} = (\dot{x}_{sat\,j,k} - X_k^-(2)) \cdot a^-_{x,j} + \cdots$$
$$\cdots + (\dot{y}_{sat\,j,k} - X_k^-(4)) \cdot a^-_{y,j} + \cdots \quad (27)$$
$$\cdots + (\dot{z}_{sat\,j,k} - X_k^-(6)) \cdot a^-_{z,j} + X_k^-(9)$$

where: $(x_{sat\,j,k}, y_{sat\,j,k}, z_{sat\,j,k})$ and $(\dot{x}_{sat\,j,k}, \dot{y}_{sat\,j,k}, \dot{z}_{sat\,j,k})$ denote the 3D position and velocity vector, respectively, of the $j^{th}$ satellite that are obtained from the ephemerides data and expressed in Cartesian coordinates; while $(X_k^-(1), X_k^-(3), X_k^-(5))$ and $(X_k^-(1), X_k^-(3), X_k^-(5))$ refer to the predicted user's absolute position and velocity vectors along the X, Y and Z axes; while $(X_k^-(7), X_k^-(8))$ are the user's clock predicted bias w.r.t to the GPS and Galileo time and $X_k^-(9)$ denotes the clock drift predictions from the EKF navigation filter.

The line-of-sight (LOS) unit vectors from the receiver to the $j^{th}$ satellite along the $X$, $Y$ and $Z$ axes are computed as follows:

$$a^-_{x,j} = \frac{(x_{sat\,j,k} - X_k^-(1))}{R^-_{j,k}}$$

$$a^-_{y,j} = \frac{(y_{sat\,j,k} - X_k^-(3))}{R^-_{j,k}} \quad (28)$$

$$a^-_{z,j} = \frac{(z_{sat\,j,k} - X_k^-(5))}{R^-_{j,k}}$$

From the pseudorange rate expression given in Eq. 19, let us denote by $V^-_{j,k}$ the relative satellite-receiver velocities without taking into account the clock drift component as:

$$V^-_{j,k} = (\dot{x}_{sat\,j,k} - x_k^-(2)) \cdot a^-_{x,j} + \cdots$$
$$\cdots + (\dot{y}_{sat\,j,k} - x_k^-(4)) \cdot a^-_{y,j} + \cdots \quad (29)$$
$$\cdots + (\dot{z}_{sat\,j,k} - x_k^-(6)) \cdot a^-_{z,j}$$

### B. CONSTRUCTION OF THE OBSERVATION MATRIX $R_k$

The predicted measurements, incorporated in the predicted measurement vector $z_k^-$, are communicated to the main EKF filter as a function of the predicted state vector $X_k^-$ through the observation (design) matrix $H_k$:

$$H_k(x_k^-) = \left.\frac{\partial h(X_k^-)}{\partial X}\right|_{X=X_k^-} \quad (30)$$

The entry $(j, m)$ of the observation matrix $H_k(j, m)$ is the partial derivative w.r.t the predicted position $\partial h_j(X_k^-)/\partial X_m$, where $j = 1 \cdots M$ refers to the $M$ tracking channels and $m = 1 \div 9$ denotes the nine states of the predicted state vector $X_k^-$.

Let us first compute the 1st row of the design matrix $H_k(j, m)$ that are the partial derivatives of the $H_k(j, m)$ entries related to the predicted pseudorange measurements $\rho^-_{j,k}$ w.r.t the predicted state vector elements $X_k^-$:

$$\left[\frac{\partial h_j\left(\rho^-_{j-GPS,k}\big|X_k^-\right)}{\partial X_k^-(1)} \cdots \frac{\partial h_j\left(\rho^-_{j-GPS,k}\big|X_k^-\right)}{\partial X_k^-(9)}\right]$$
$$= \cdots$$
$$\cdots = [-a^-_{x,j}\; 0\; -a^-_{y,j}\; 0\; -a^-_{z,j}\; 0\; 1\; 0\; 0]^T_{N_{GPS}\times 1} \quad (31)$$

$$\left[\frac{\partial h_j\left(\rho^-_{j-GAL,k}\big|X_k^-\right)}{\partial X_k^-(1)} \cdots \frac{\partial h_j\left(\rho^-_{j-GAL,k}\big|X_k^-\right)}{\partial X_k^-(9)}\right]$$
$$= \cdots$$
$$\cdots = [-a^-_{x,j}\; 0\; -a^-_{y,j}\; 0\; -a^-_{z,j}\; 0\; 1\; 0]^T_{(M-N_{GPS})\times 1}$$

where: $\rho^-_{j-GPS,k}$ and $\rho^-_{j-GAL,k}$ represent the predicted pseudoranges to the The GPS and Galieleo satellites, respectively and $N_{GPS}$ denotes the number of tracked GPS satellites.

The remaining M to 2M rows of the design matrix $H_k(j,m)$ include the partial derivatives of the predicted pseudorange rates measurements $\dot\rho^-_{j,k}$ w.r.t the predicted state vector $X^-_k$. The partial derivatives are computed separately for the position and velocity terms of the predicted state vector $X^-_k$. Regarding the X-position related terms, the following relations can be written:

$$v^-_{x,j} = \frac{\partial h_j\left(\dot\rho^-_{j,k}\middle| X^-_k\right)}{\partial X^-_k(1)} = \left(x_{sat\,j,k} - X^-_k(1)\right) \cdot \\ \cdots \cdot \frac{V^-_{j,k}}{{R^-_{j,k}}^2} - \frac{\left(\dot x_{sat\,j,k} - X^-_k(2)\right)}{R^-_{j,k}} \quad (32)$$

Similarly, for the partial derivatives of the pseudorange rates w.r.t the predicted user position along the Y and Z-axes, denoted respectively as $X^-_k(3)$ and $X^-_k(5)$:

$$v^-_{y,j} = \frac{\partial h_j\left(\dot\rho^-_{j,k}\middle| x^-_k\right)}{\partial X^-_k(3)} = \left(x_{sat\,j,k} - X^-_k(3)\right) \cdot \\ \cdots \cdot \frac{V^-_{j,k}}{{R^-_{j,k}}^2} - \frac{\left(\dot x_{sat\,j,k} - X^-_k(4)\right)}{R^-_{j,k}}$$

$$v^-_{z,j} = \frac{\partial h_j\left(\dot\rho^-_{j,k}\middle| x^-_k\right)}{\partial X^-_k(5)} = \left(x_{sat\,j,k} - X^-_k(5)\right) \cdot \\ \cdots \cdot \frac{V^-_{j,k}}{{R^-_{j,k}}^2} - \frac{\left(\dot x_{sat\,j,k} - X^-_k(6)\right)}{R^-_{j,k}} \quad (33)$$

On the other side, the design matrix $H_k(j,m)$ elements corresponding to the partial derivatives of the predicted pseudorange rates measurements $\dot\rho^-_{j,k}$ w.r.t the velocity terms of the predicted state vector $X^-_k$, are computed as follows:

$$v^-_{\dot x,j} = \frac{\partial h_j\left(\dot\rho^-_{j,k}\middle| X^-_k\right)}{\partial X^-_k(2)} = -a^-_{x,j}$$

$$v^-_{\dot y,j} = \frac{\partial h_j\left(\dot\rho^-_{j,k}\middle| X^-_k\right)}{\partial X^-_k(4)} = -a^-_{y,j} \quad (34)$$

$$v^-_{\dot z,j} = \frac{\partial h_j\left(\dot\rho^-_{j,k}\middle| X^-_k\right)}{\partial X^-_k(6)} = -a^-_{z,j}$$

and w.r.t the clock bias drift term $x_7$ of the predicted state vector $X^-_k$:

$$(35)$$

$$v^-_{c \cdot \dot t_{clk}, j, k} = \frac{\partial h_j\left(\dot\rho^-_{j,k}\middle| X^-_k\right)}{\partial X^-_k(9)} = 1$$

Finally, the observation matrix $H_k$ that is required for the Kalman gain computation, is given in Eq. (36):

$$H_k = \begin{bmatrix} -a^-_{x,1} & 0 & -a^-_{y,1} & 0 & -a^-_{z,1} & 0 & 1 & 0 \\ -a^-_{x,2} & 0 & -a^-_{y,2} & 0 & -a^-_{z,2} & 0 & 1 & 0 \\ \vdots & \vdots & \vdots & 0 & \vdots & \vdots & \vdots & \vdots \\ -a^-_{x,M} & 0 & -a^-_{y,2} & 0 & -a^-_{z,2} & 0 & 1 & 0 \\ v^-_{\dot x,1} & -a^-_{x,1} & v^-_{\dot y,1} & -a^-_{y,1} & v^-_{\dot z,1} & -a^-_{z,1} & 0 & 1 \\ v^-_{\dot x,2} & -a^-_{x,2} & v^-_{\dot y,2} & -a^-_{y,2} & v^-_{\dot z,2} & -a^-_{z,2} & 0 & 1 \\ \vdots & \vdots & \vdots & \vdots & \vdots & \vdots & \vdots & \vdots \\ v^-_{\dot x,M} & -a^-_{x,M} & v^-_{\dot y,M} & -a^-_{y,M} & v^-_{\dot z,M} & -a^-_{z,M} & 0 & 1 \end{bmatrix}_{2M \times 8} \quad (36)$$

## C. EKF INNOVATION VECTOR

The code delay and frequency carrier estimation process are achieved per channel basis as in the scalar configuration, however in the vectorized architecture, the DLL and FLL discriminator outputs will be fed to the EKF navigation filter as its measurement innovation vector. The state vector estimate update $X^+_k$ is obtained using the following expression:

$$X^+_k = X^-_k + K_k \cdot \delta z_k = X^-_k + K_k \cdot \delta z_k \quad (37)$$

Where $\delta z_k$ represents the measurement innovation vector, including the pseudorange and pseudorange rate errors for each tracking channel $j = 1 \div M$ that are computed from the DLL and FLL discriminator outputs, is given as:

$$\delta z_k = z(k) - h(X^-_k) \\ = [\delta\rho_1, \delta\rho_2 \cdots \delta\rho_M \;\vdots\; \delta\dot\rho_1, \delta\dot\rho_2 \cdots \delta\dot\rho_M\,]_k \quad (38)$$

where the first $M$ terms, related to the pseudorange errors, are computed from the DLL discriminator outputs using the following relation:

$$\delta z_{\delta\rho|k} = \left[\left(\frac{c}{f_{code}}\right) \cdot \left(D_{DLL,1}, \cdots, D_{DLL,M}\right)\right]_k \quad (39)$$

Similarly, the pseudorange rate errors computation for each channel is achieved from the FLL discriminator outputs:

$$\delta z_{\delta\dot\rho_{FLL}|k} = \left[\left(\frac{c}{f_{carr}}\right) \cdot \left(D_{FLL,1}, \cdots, D_{FLL,M}\right)\right]_k \quad (40)$$

## IV. VDFLL FEEDBACK LOOP: CODE AND CARRIER NCO UPDATE

The code and carrier NCO update is performed per each tracked channel $j$ based on the EKF state vector prediction $X^-_k$ from Eq. (4). The pseudorange rate prediction, including the contribution of the satellite clock drift error $\dot b_{sv-c,j,k}$, is given by:

$$\dot{\rho}^-{}_{j,k+1} = \left(\dot{x}_{sat\ j,k+1} - X^-_{k+1}(2)\right) \cdot a^-_{x,j} + \cdots$$
$$\cdots + \left(\dot{y}_{sat\ j,k+1} - X^-_{k+1}(4)\right) \cdot a^-_{y,j} + \cdots$$
$$\cdots + \left(\dot{z}_{sat\ j,k+1} - X^-_{k+1}(6)\right) \cdot a^-_{z,j} + X^-_{k+1}(9) \quad (41)$$
$$+ \dot{b}_{sv-c,j,k+1}$$

The Doppler frequency correction $\delta f^-_{D_{j,k+1}}$ per each tracking channel $j$, closing the feedback loop to the carrier NCO, is computed by projecting the predicted velocity- and clock drift errors states in the pseudorange rate error domain as:

$$f_{NCO-ca,j,k+1} = \delta f^-_{D_{j,k+1}} = \frac{f_{carr}}{c} \cdot \dot{\rho}^-_{j,k+1}\ (Hz) \quad (42)$$

where: $f_{carr} = 1{,}57542\ GHz$ refers to GPS L1 & Galileo E1 carrier frequency and $c = 3 \cdot 10^8$ is the speed of light in (m/s).

On the other hand, the code NCO command for each channel $j$ is forwarded to successive tracking epoch by taking the difference between the pseudorange predictions of two consecutive measurement epochs, denoted as $\rho^-{}_{j,k+1}$ and $\rho^-{}_{j,k}$, respectively:

$$f_{NCO-co,j,k+1} = f_{code} \cdot \frac{(\rho^-{}_{j,k+1} - \rho^-{}_{j,k})}{c \cdot T_{EKF}} \quad (43)$$

where $T_{EKF}$ is the EKF update time set to the code and carrier accumulation period.

## V. Performed tests

In order to test the performance of the proposed L1/E1 VDFLL architecture, a GNSS emulator compiled in C language, able to generate GPS L1 and Galileo E1 signals up to 48 channels simultaneously, was used. Moreover, the vector tracking algorithm is implemented in C language platform, driven by the faster execution time of KF algorithm at high rates (set equal to the tracking outputs at 50 Hz, $T_{EKF} = 20ms$). Three distinctive GNSS receiver architectures will be analyzed with the scope of performance comparison:

- Scalar tracking employing a 3rd order loop PLL and a DLL, with a KF positioning module at 1 Hz for the PVT computation, where the pseudorange and Doppler measurements are included in the observation vector.
- The same scalar tracking architecture but now integrated with a KF positioning module at 50 Hz, similar to the VDFLL algorithm update rate.
- The proposed VDFLL EKF architecture working at $T_{EKF} = 20ms$ integration time and thus providing 50 Hz code and carrier frequency updates.

It must be noted the KF positioning module is similar to the EKF filter of the vectorized solution, with the differences that a closed-loop measurement covariance matrix is used in the former and moreover, the KF filter operates on locked satellites only whereas the VDFLL uses all satellites in view.

The simulations performed in this work are related to two different types of user trajectories: first, a static user and secondly, a real car trajectory in Toulouse urban area. The simulated reception conditions will consist in several signal outages and significant power drops simulated in different satellite channels in order to observe the tracking performance of the proposed VDFLL architecture with respect to conventional tracking. In both test scenarios, there is maximum of 13 simultaneously tracked GPS L1 and Galileo E1 channels during 200 GPS epochs.

A detailed performance comparison between the scalar and vectorized configurations will be assessed in two different levels:

- *System level*: expressed in terms of user's position and velocity estimation accuracies, position and velocity errors statistics and resistance to degraded signal reception conditions;

- *Channel level*: indicated by the code delay and carrier Doppler frequency estimation errors and their standard deviations in the presence of outages.

In details, the code and carrier tracking parameters used by the scalar configuration and the vectorized architecture are summarized in Table 1.

TABLE I.  CODE AND CARRIER TRACKING PARAMETERS EMPLOYED IN THE SCALAR AND VECTORIZED ARCHITECTURES

| L1/E1 Code Tracking Parameters | |
|---|---|
| DLL order | 1 |
| DLL noise bandwidth ($B_{DLL-n}$) | 1 Hz |
| DLL period | 0.02 s |
| Code delay discriminator | Early Minus Late Power (EMLP) |
| GPS L1 chip spacing ($k_{cs-L1}$) | 0.5 chips |
| GAL E1 chip spacing ($k_{cs-E1}$) | 0.2 chips |
| L1/E1 Carrier Tracking Parameters | |
| *Scalar Configuration* | |
| PLL order | 3 |
| PLL noise bandwidth ($B_{PLL-n}$) | 10 Hz |
| PLL period | 0.02 s |
| Carrier phase discriminator | Costas Discriminator |
| *Vectorized Architecture* | |
| Carrier frequency period | 0.01 s |
| Carrier frequency discriminator | Cross Dot Product |

The simulations herein presented use the GPS and Galileo constellations in the L1 band, taking into consideration the binary phase shift keying BPSK(1) modulation for GPS L1 and the binary offset carrier modulation BOC(1,1) for Galileo E1. It must be noted that a detector lock and that a hot start re-acquisition process of 1s for both the scalar tracking configurations are implemented. Moreover, the initial assumed code and carrier estimation errors are modeled as zero-mean Gaussian noise terms with standard deviations set according to the code chip spacing and carrier phase period, respectively.

As previously stated, the received signals were simulated at the correlator output level in an ENAC-owned semi-analytic receiver simulator. An RF front-end with a 24 MHz bandwidth (double-sided) is assumed. Multiple outages on different tracked satellites have been simulated, by generating a sudden drop of the $CN_0$ ratio down to 20 dB-Hz that coincides with the $CN_0$ level in quasi-indoor environment [9]. Therefore, the GNSS signals exhibiting this low $CN_0$ level will cause the tracking loops to experience a loss of lock condition. The outage conditions were simulated in three time epochs as depicted in Fig. 3:

1. Outage 1 from the $2^{nd} - 12^{th}$ time epoch (10 seconds);
2. Outage 2 from the $60^{th} - 80^{th}$ time epoch (20 seconds);
3. Outage 3 from the $140^{th} - 160^{th}$ time epoch (20 seconds);

The oscillator's phase noise PSD $\sigma_b$ and the oscillator's frequency noise PSD $\sigma_d$, which by themselves depend on the Allan variance parameters $h_0$ and $h_{-2}$ [2], are given as:

$$S_{c\varphi} = \omega_c^2 \cdot \frac{h_0}{2}$$
$$S_{cf} = 2\pi^2 \cdot \omega_c^2 \cdot h_{-2}$$
(44)

In our implementation, a Temperature Controlled Oscillator (TCXO) is chosen, where $\omega_c = 2\pi \cdot f_{carr}$ is the carrier frequency expressed in radians and the white noise frequency ($h_0$) and integrated frequency noise ($h_{-2}$) have the following values:
$$h_0 = 1 \cdot 10^{-21}$$
$$h_{-2} = 2 \cdot 10^{-20}$$
(44)

In the following subsections, the performance comparison between the conventional tracking and the vectorized algorithm will be exploited firstly for the static user case and afterwards, for the car trajectory scenario.

**(A) Static user**

In the fixed user scenario the simulated receiver is located at the Signal and Navigation (SIGNAV) laboratory premises in ENAC, Toulouse with coordinates 43° 33' 56.688" N, 1° 28' 49.796" W and altitude 200 m. The performance comparison between the VDFLL architecture and the two scalar tracking configurations will be examined in details under signal outages conditions.

For this matter, three intervals of signal's $CN_0$ drops were introduced for two GPS L1 and Gal E1 satellites, specifically GPS PRN 3 & 4 and Galileo PRN 51 & 52, as plotted in Fig. 4. While the performance analysis of the scalar tracking algorithms and VDFLL algorithm in the position error and code delay discriminator error domain is illustrated in the second and third subplots of Fig. 4. The position error along the three ECEF axes and the clock bias estimation error from the EKF filter is given in details for the two architectures for both the outages and no outages epochs in Table II. For the scalar tracking receiver, the presence of signal blockages creates a loss of lock on the affected satellites. Thus, the pseudorange and Doppler information of the four satellites under outage are not passed to the EKF navigation processor. Whereas, no satellite lock test is performed in the VDFLL architecture, where the observations from all the satellites in view are fed to the EKF estimation filter.

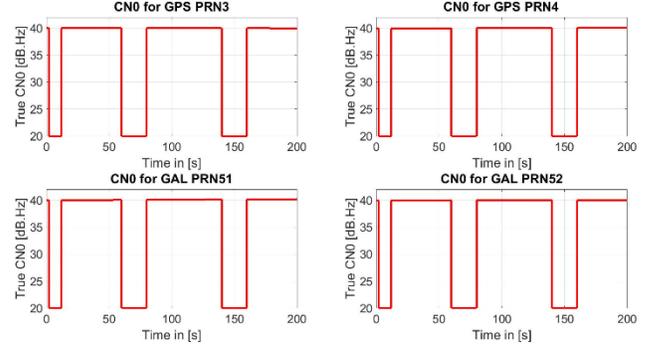

**Fig. 3.** Illustration of CN0 ratio drops to 20 dB-Hz for four satellite namely, GPS PRN 3 & 4 and GAL PRN 51 & 52 during three outages intervals.

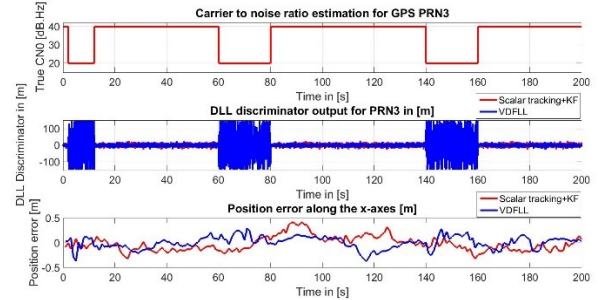

**Fig. 4.** Performance comparison between the scalar tracking @ 1 Hz rate (in red) and VDFLL algorithm (in blue) for a static user under signal outages for GPS PRN 3 in terms of: **a)** CN0 estimation in dB-Hz; **b)** DLL discriminator output in [m]; **c)** Position error along the X-axis in [m].

Similar position and clock bias error bounds can be observed for the two architectures under outages conditions, as can be seen in Table II. The reasons are twofold: On the first place, due to the overdetermined number of observations fed to the EKF navigation filter, specifically 2x9 pseudoranges and Doppler measurements for the scalar configuration and 2x13 for the vectorized architecture and on the second place, related to the high $CN_0$ reception of the tracked satellites.

The performance analysis in the signal level, in terms of code delay and Doppler frequency estimation errors is illustrated in Fig. 5 and 6. During the three outage intervals, as shown in Fig. 5.a), the VDFLL is able to predict the carrier and code parameters by exploiting its EKF prediction model even though the discriminator error for the affected channel reaches maximum levels.

TABLE II. POSITION AND CLOCK BIAS ESTIMATION ERROR STATISTICS FOR THE SCALAR TRACKING + KF MODULE AND THE PROPOSED VDFLL ARCHITECTURE

|  | Scalar + KF | | VDFLL | |
|---|---|---|---|---|
|  | **E[m]** | **Std[m]** | **E[m]** | **Std[m]** |
| *No outage* | | | | |
| **X-error** | 0.137 | 0.11 | 0.011 | 0.13 |
| **Y-error** | 0.061 | 0.077 | 0.007 | 0.114 |
| **Z-error** | 0.214 | 0.167 | 0.048 | 0.017 |
| **Clock bias error** | 0.216 | 0.436 | 0.003 | 0.13 |
| | | | | |
| *Outage* | | | | |
| **X-error** | 0.160 | 0.18 | 0.17 | 0.14 |
| **Y-error** | 0.122 | 0.001 | 0.03 | 0.167 |
| **Z-error** | 0.223 | 0.037 | 0.20 | 0.219 |
| **Clock bias error** | 0.251 | 0.176 | 0.17 | 0.148 |
| *Comment :* E[] - mean value  Std[] - standard deviation | | | | |

The traditional tracking does not rely on any model to propagate the code and carrier estimations, but instead on current observations only. Therefore, a reacquisition process is initiated directly after the end of the outage period, with the objective of re-locking the previously "lost" signal, as is illustrated in red in the zoomed plot during the second outage period in Fig 5 b). The straight red line in Fig 5 b), clearly states that in conventional tracking no code delay NCO update is computed during the outage period for PRN3 since the lock detection test is not passed. Directly after, a 1 second reacquisition procedure is initiated, from the 80th epoch as illustrated in Fig 5 b), aiming to the provision of a rough code delay estimation. On the contrary, the VDFLL tracking mechanism significantly improves the code/carrier tracking function without the requirement of a reacquisition process but only relying on its EKF prediction model.

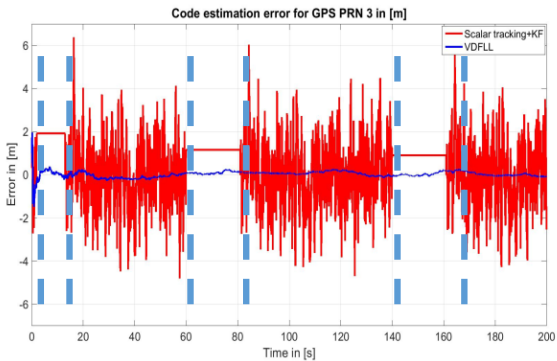

a)

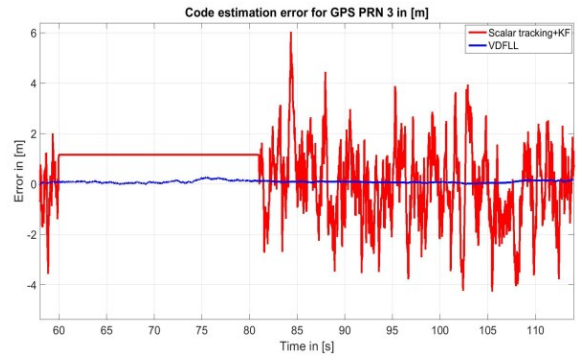

b)

**Fig. 5. a)** The GPS PRN 3 code delay estimation error (in m) for the scalar tracking (in red) and VDFLL algorithm (in blue) for a static user under outage conditions (in blue dashed lines); **b)** Zoomed view of the code estimation error of a) focusing in the 1 second reacquisition process starting directly after the end of the 2nd outage at the 80th epoch.

The same logic holds even for the Doppler frequency estimation in Fig 6. The code delay and Doppler frequency estimation statistics in terms of their mean value and standard deviation for the scalar and vectorized architectures are summarized in Table III.

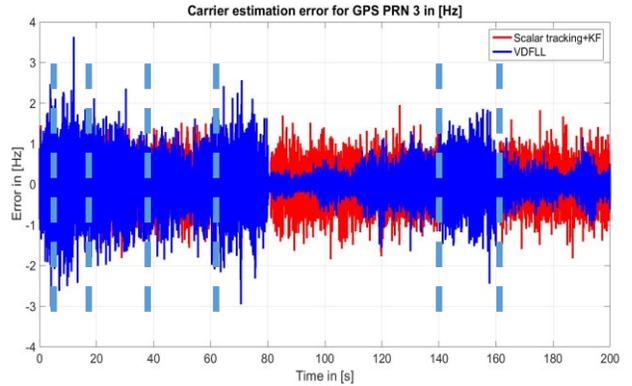

**Fig. 6.** The Doppler frequency estimation error (in Hz) for the scalar tracking (in red) and VDFLL algorithm (in blue) for a static user under outage conditions (in dashed blue lines) for GPS PRN 3.

TABLE III. CODE DELAY AND CARRIER DOPPLER FREQUENCY ESTIMATION STATISTICS FOR THE SCALAR TRACKING + KF MODULE AND THE PROPOSED VDFLL ARCHITECTURE

|  | Scalar + KF | | VDFLL | |
|---|---|---|---|---|
|  | **E[]** | **Std[]** | **E[]** | **Std[]** |
| *No outage* | | | | |
| **Code estimation error (m)** | 0.067 | 1.489 | 0.079 | 0.08 |
| **Doppler freq. estimation error (Hz)** | 0.007 | 0.49 | 0.035 | 0.30 |
| | | | | |
| *Outage* | | | | |
| **Code estimation error (m)** | - | - | 0.1 | 0.11 |

| Doppler freq. Estimation (Hz) | - | - | 0.15 | 0.747 |
|---|---|---|---|---|
| *Comment :*  E[]  - mean value  Std[] - standard deviation | | | | |

A marked degradation of the scalar tracking concerning the code/carrier tracking standard deviation values can be easily noticed. In specifics, lower values of the code/Doppler estimations are observed for the L1/E1 VDFLL architecture, especially concerning the code delay estimation during normal operation. The likely reason for this behavior is linked to the inter-channel aiding through the common EKF estimation filter and the estimation update based on the forward position/velocity projection in the vectorized architecture.

**(B) Dynamic user**

In particular, a realistic car trajectory in high dynamic condition is generated based on the reference trajectory computed by the NovAtel's SPAN receiver mounted on car during a 40 minutes measurement campaign conducted in Toulouse. It must be noted that the simulated car path of 200 seconds duration is a representative of a car trajectory but not of urban signal reception conditions since the simulated received conditions are generated from an open sky environment plus the forced drops of received signal $C/N_0$ values. Besides, since the reference trajectory is output at 1 Hz rate, interpolation is used to generate the true trajectory at the VDFLL EKF filter rate @ 50 Hz. The simulated car path is shown in Fig. 7. The outage scenario, shown in Fig. 4, is also applied to the dynamic user case. Moreover, the position domain comparison is extended to the scalar tracking architecture with the KF module working in the same rate as the vectorized architecture that is 50 Hz. The position error plots in the ECEF frame illustrated in Fig. 8, demonstrate a clear convergence of the VDFLL-computed navigation solution to the reference trajectory within a standard deviation of 0.2 m during the outages, as illustrated in Table IV.

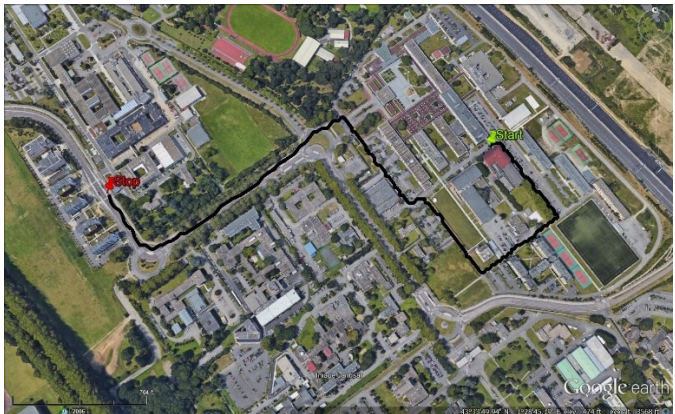

**Fig. 7.** The reference trajectory along the city of Toulouse, France.

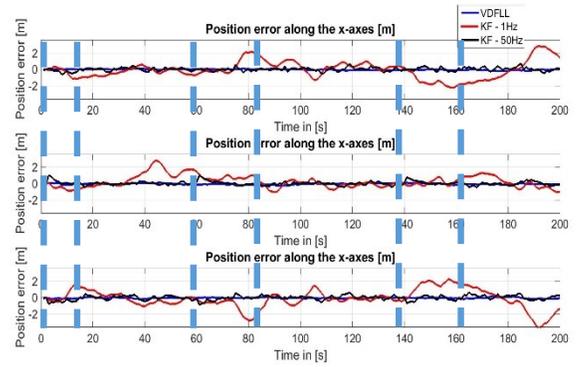

**Fig. 8.** Position error comparison between the scalar tracking + KF positioning module @ 1 Hz (in red), the VDFLL algorithm (in blue) and the scalar tracking + KF positioning module @ 50 Hz (in black) for a car trajectory under signal outages (in dashed aqua line) along the: **a)** X-axis in [m]; **b)** Y-axis in [m]; **c)** Z-axis in [m]

TABLE IV. POSITION AND CLOCK BIAS ESTIMATION ERROR STATISTICS FOR THE SCALAR TRACKING + KF MODULE AND THE PROPOSED VDFLL ARCHITECTURE

|  | Scalar + KF Scalar @ 1 Hz | | VDFLL | |
|---|---|---|---|---|
|  | E[m] | Std[m] | E[m] | Std[m] |
|  | *No outage* | | | |
| **X-error** | 0.507 | 0.72 | 0.03 | 0.06 |
| **Y-error** | 0.07 | 0.458 | 0.044 | 0.10 |
| **Z-error** | -0.289 | 0.818 | 0.05 | 0.058 |
| **Clock bias error** | 0.17 | 0.401 | 0.06 | 0.047 |
|  | Scalar + KF @ 50 Hz | | | |
|  | E[m] | Std[m] | | |
|  | *No outage* | | | |
| **X-error** | 0.12 | 0.22 | | |
| **Y-error** | 0.05 | 0.184 | | |
| **Z-error** | 0.06 | 0.211 | | |
| **Clock bias error** | 0.07 | 0.132 | | |
|  | *Outage* | | | |
| **X-error** | 0.199 | 0.874 | -0.06 | 0.16 |
| **Y-error** | 0.707 | 0.715 | -0.087 | 0.204 |
| **Z-error** | -0.884 | 0.777 | 0.026 | 0.19 |
| **Clock bias error** | -0.09 | 0.46 | 0.019 | 0.197 |
|  | Scalar + KF @ 50 Hz | | | |
|  | E[m] | Std[m] | | |
|  | *Outage* | | | |
| **X-error** | 0.08 | 0.19 | | |
| **Y-error** | - 0.11 | 0.315 | | |
| **Z-error** | - 0.06 | 0.327 | | |
| **Clock bias error** | 0.02 | 0.22 | | |
| *Comment :*  E[]  - mean value  Std[] - standard deviation | | | | |

The positioning root mean square errors (RMSE) per each ECEF axes illustrated in Fig. 8, highlight the positioning robustness of the vectorized architecture during both outages and high $CN_0$ reception conditions. It is obvious the positioning error variance increase of the scalar tracking + KF module operating at 1 Hz during the outages intervals, with an error peak up to 2.75 m at the 80th time epoch. Moreover, a sudden navigation error increase is also observed at the 190th epoch coming from the last car turn prior to arriving at the end of the path in Fig. 7. Concerning the proposed VDFLL algorithm, a smoother position error in the ECEF frame is marked along the overall car trajectory, with significantly smaller errors. An interesting result is observed for the scalar architecture employing the KF positioning module at 50 Hz rate that clearly outperforms its counterpart running at 1 Hz, in terms of position error bounding. The explanation lies on the higher rate of the state propagation and measurement innovation process, providing a faster convergence of the position estimations to the reference trajectory. Nevertheless, quasi similar positioning errors are observed w.r.t to the VDFLL architecture and this comes due the overdetermined number of observables (13 tracked satellite channels in total).

In the signal tracking level, the code delay and Doppler frequency estimation plots follow the same trend as previously illustrated for the static scenario in Fig. 5 and 6. However, clear insight of the code/carrier estimation statistics for the dynamic case are given in Table V. In contrast to the scalar tracking, the proposed vectorized architecture is capable of continuously estimating the code delay/carrier Doppler of the outage-affected satellite. However, it must be noted that these estimations exhibit a higher noise level due to the fact that the "corrupted" satellite measurements are still fed to the EKF estimation filter.

TABLE V. CODE DELAY AND CARRIER DOPPLER FREQUENCY ESTIMATION STATISTICS FOR THE SCALAR TRACKING + KF MODULE AND THE PROPOSED VDFLL ARCHITECTURE

|  | Scalar + KF | | VDFLL | |
|---|---|---|---|---|
|  | E[] | Std[] | E[] | Std[] |
| *No outage* | | | | |
| Code estimation error (m) | 0.07 | 1.50 | 0.079 | 0.08 |
| Doppler freq. estimation error (Hz) | 0.007 | 0.52 | 0.03 | 0.54 |
|  | | | | |
| *Outage* | | | | |
| Code estimation error (m) | - | - | 0.05 | 0.15 |
| Doppler freq. Estimation (Hz) | - | - | 0.21 | 0.78 |
| Comment :  E[] - mean value  Std[] - standard deviation | | | | |

In order to fully exploit the VDFLL capabilities in dynamic scenarios under bad signal reception conditions, the navigation solution has been computed with only 3 satellites in view, which is harsher than the minimum requirement for a conventional Least Square (LS) PVT computation. In other words, the navigation solution for the two scalar architectures is computed by using the three locked satellites only, whereas in the VDFLL algorithm all the six tracked satellites (comprising the three "corrupted" ones) are used. The other satellite channels are in outage at the same periods as in the previous tests. Obviously, due to the poor availability of visible satellites and their bad geometric distribution, the two scalar tracking + KF navigation (operating at 1 Hz and 50 Hz) computed trajectories, illustrated in red and black, respectively, diverge from the reference one during the outages, as shown in Figure 10. This divergence is strictly related to the scalar tracking + KF position module operation mode that uses only the observations coming from the locked satellite channels and clearly 3 measurements are not sufficiently enough for a correct state vector estimation. An interesting result consists on the worse positioning performance of the KF @ 50 Hz (illustrated in green) w.r.t its 1 Hz operating counterpart (in red) under outages. We believe that the reason of this deterioration, exhibiting position error peaks of nearly 23 m is due its high operation rate which in KF estimation terms is translated into an increased trust on the state prediction (lower process noise Q matrix terms). On the other hand, the vectorized navigation solution is less oscillating and clearly follows the reference trajectory within a mismatching value up to 2.7 m during the signals $CN_0$ drops. The explanation lies on the higher numberr of observations fed to the navigation filters that works with tracked and not locked satellites and moreover, on the "corrupted" measurement de-weighting procedure through the measurement covariance matrix $R_k$. The 3D positioning performance comparison between the three tested architectures w.r.t to the reference trajectory is given in Earth North Up (ENU) coordinates in Fig. 11.

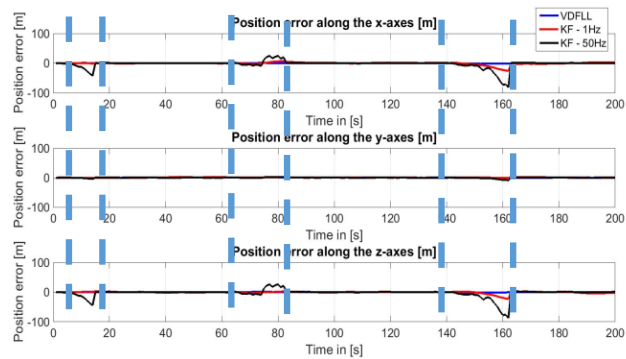

**Fig. 10.** Position error comparison between the scalar tracking + KF positioning module @ 1 Hz (in red), the VDFLL algorithm (in blue) and the scalar tracking + KF positioning module @ 50 Hz (in black) for a car trajectory under signal outages (in dashed aqua line) for 3 visible satellites only along the: **a)** X-axis in [m]; **b)** Y-axis in [m]; **c)** Z-axis in [m].

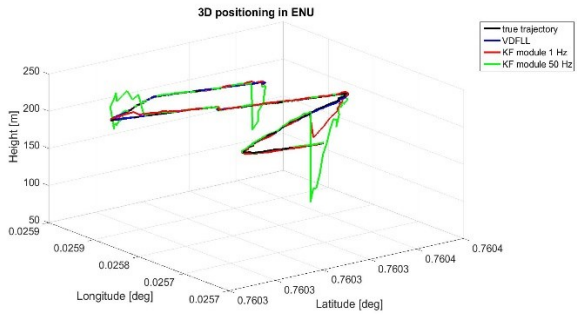

**Fig. 11.** 3D Navigation solution comparison in ENU frame between the scalar tracking + KF positioning module @ 1 Hz (in red), the VDFLL algorithm (in blue) and the scalar tracking + KF positioning module @ 50 Hz (in green) for a car trajectory (in black) under signal outages (in dashed aqua line) for 3 visible satellites only along the: **a)** X-axis in [m]; **b)** Y-axis in [m]; **c)** Z-axis in [m].

The position error statistics of the proposed L1/E1 VDFLL architecture in the extreme case of only 3 satellites in view during the outages, are summarized in Table VI.

TABLE VI. POSITION AND CLOCK BIAS ESTIMATION ERROR STATISTICS FOR THE PROPOSED VDFLL ARCHITECTURE WITH ONLY 3 VISIBLE SATELLITES

|  | VDFLL | | | |
|---|---|---|---|---|
|  | *No outage* | | *Outage* | |
|  | **E[m]** | **Std[m]** | **E[m]** | **Std[m]** |
| **X-error** | -0.11 | 0.14 | -1.15 | 0.622 |
| **Y-error** | 0.06 | 0.07 | 0.1 | 0.14 |
| **Z-error** | 0.15 | 0.08 | -1.26 | 0.623 |
| **Clock bias error** | -0.73 | 0.1 | -0.82 | 0.28 |
| *Comment :* E[] - mean value  Std[] - standard deviation | | | | |

## VI. CONCLUSIONS AND FUTURE WORK

In this paper, a vector delay/frequency-locked loop (VDFLL) architecture for a dual constellation L1/E1 GPS/Galileo receiver is proposed. After the mathematical description of the EKF filter's prediction and observation model, a detailed performance comparison in the position and tracking domain between the scalar tracking + KF positioning operating at two different rates (1 and 50 Hz) and VDFLL configuration was assessed under signal outages simulated in different satellite channels. The results for both the static and dynamic scenarios showed that contrary to the conventional tracking, the L1/E1 VDFLL loop is able to recover the frequency and code-delay estimation at the end of the simulated outages without the requirement of signal reacquisition process. Moreover, VDFLL provides better performance, both in the position domain and the tracking level, than the scalar architecture especially in the dynamic scenarios for a reduced number of satellites in view. However, in high number of observations scenario, there is no real gain of employing the VDFLL architecture and instead only an increased update rate of the EKF filter in the scalar tracking configuration is sufficient.

The likely reason for this behavior is linked to the inter-channel aiding through the update process based on the forward position/velocity projection in the vectorized architecture. The full capability of the L1/E1 VDFLL architecture was exploited in the last performed test, where only three satellites were in view during outage conditions. Even in this harsh case, the vectorized algorithm was not only able of providing an available navigation solution but also assuring an accurate estimation within the 3 m position error bound.

Future work will proceed on three fronts. First, the detailed performance analysis concerning the position and tracking accuracy will be extended to the presence of multipath-simulated signals, generated by the DLR Land Mobile Channel Model (LMCM) that will be adapted to the multi-channel tracking mechanism. Second, more testing will be performed on the already-described L1/E1 VDFLL architecture and the L1/E1 EKF tracking per channel with the vectorized navigation filter proposed for an increased tracking capability, both able to work in vector-tracking mode. Last but not least, the vectorized architecture will be extended to the carrier phase estimation in order to fully accomplish the positioning and tracking capability of vector tracking in signal-constrained environment.

ACKNOWLEDGMENT

This work was financially supported by EU FP7 Marie Curie Initial Training Network MULTI-POS (Multi-technology Positioning Professionals) under grant nr. 316528.